\documentclass[nofootinbib,prd,twocolumn,showpacs,showkeys,preprintnumbers]{revtex4-1}
\usepackage{hyperref,amssymb,amsmath,mathrsfs,bm,graphicx,stackrel}
\begin{document}
\title {The spacetime outside a source of gravitational radiation: The axially symmetric null fluid }
\author{L. Herrera}
\email{lherrera@usal.es}
\affiliation{Escuela de F\'\i sica, Facultad de Ciencias, Universidad Central de Venezuela, Caracas, Venezuela and Instituto Universitario de F\'isica
Fundamental y Matem\'aticas, Universidad de Salamanca, Salamanca, Spain}
\author{A. Di Prisco}
\email{alicia.diprisco@ciens.ucv.ve}
\affiliation{Escuela de F\'\i sica, Facultad de Ciencias, Universidad Central de Venezuela, Caracas, Venezuela}
\author{J. Ospino}
\email{j.ospino@usal.es}
\affiliation{Departamento de Matem\'atica Aplicada and Instituto Universitario de F\'isica
Fundamental y Matem\'aticas, Universidad de Salamanca, Salamanca, Spain}
\begin{abstract}
We carry out a study  of the exterior of an   axially and reflection symmetric source of gravitational radiation.  The exterior of such a source is filled with   a null fluid produced by the dissipative processes inherent to the emission of gravitational radiation, thereby representing a generalization of the Vaidya metric  for  axially and reflection symmetric spacetimes.  The role of the vorticity, and its relationship with the  presence of gravitational radiation is put in evidence. The spherically symmetric case (Vaidya) is, asymptotically, recovered within the context of the  $1+3$ formalism. 
\end{abstract}
\maketitle

\section{Introduction}
It is known that in the hydrodynamic description of a  physically meaningful source (i.e. bounded and regular) of gravitational radiation, there should be present a dissipative  term, which accounts for the increasing of entropy associated to such an emission \cite{7}, \cite{hetaln}. Accordingly, we should  expect that any exterior of such a source should  entail the presence of incoherent radiation (null fluid), associated to those irreversible processes. 

It is our purpose in this manuscript, to provide a full description of the spacetime surrounding a  bounded source of gravitational radiation. For simplicity we shall impose the highest degree of symmetry compatible with the presence of gravitational radiation, i.e.   axially and reflection symmetry.  Thus, the exterior of such a source is filled with  a null fluid, and represents  a generalization of the Vaydia metric  for  axially and reflection symmetric spacetimes. 

In this work we shall heavily rely on the formalism developed in \cite{1}, which is based on the $1+3$ formalism \cite{21cil}--\cite{nin}, thus, even though we shall try to make this manuscript, as self--contained as possible, we shall frequently refer the reader to \cite{1},  in order to avoid the rewriting of some of the equations.

In the next section we shall provide a summary of the main equations and concepts used in this study. Next we describe the null fluid outside the source. In order to stress the role of the vorticity in the emission of gravitational radiation we shall consider the vorticity--free case, which is shown to lead to either the static case or to the spherically symmetric case (Vaidya). This latter case is analyzed in some detail. A summary of the results is presented in the last section, and some intermediate equations are deployed in  an Appendix.

\section{Basic definitions and notation}
In this section we shall deploy all the variables required for our study, some details of the calculations are given  in \cite{1}, and therefore we shall omit them here.
\subsection{The metric, the source, and the kinematical variables}

We shall consider,  axially (and reflection) symmetric space--times. For such  systems the  line element may be written in ``Weyl spherical coordinates'' as:

\begin{equation}
ds^2=-A^2 dt^2 + B^2 \left(dr^2
+r^2d\theta^2\right)+C^2d\phi^2+2Gd\theta dt, \label{1b}
\end{equation}
where $A, B, C, G$ are positive functions of $t$, $r$ and $\theta$. We number the coordinates $x^0=t, x^1=r, x^2= \theta, x^3=\phi$.

We shall assume that  our source is filled with an anisotropic and dissipative fluid, and is bounded by a timelike surface $\Sigma$, where junction (Darmois) conditions should be imposed.

The energy momentum tensor of the source may be written in the ``canonical'' form, as 
\begin{equation}
{T}_{\alpha\beta}= (\mu+P) V_\alpha V_\beta+P g _{\alpha \beta} +\Pi_{\alpha \beta}+q_\alpha V_\beta+q_\beta V_\alpha.
\label{6bis}
\end{equation}

In the above expression  $V^\mu$ denotes the four-velocity assigned by certain observer,  $\mu$ is the energy
density (the eigenvalue of $T_{\alpha\beta}$ for eigenvector $V^\alpha$), $q_\alpha$ is the  heat flux, whereas  $P$ is the isotropic pressure, and $\Pi_{\alpha \beta}$ is the anisotropic tensor. 

Since we choose the fluid to be comoving in our coordinates, then
\begin{equation}
V^\alpha =\left(\frac{1}{A},0,0,0\right); \quad  V_\alpha=\left(-A,0,\frac{G}{A},0\right).
\label{m1}
\end{equation}
Next, let us  introduce the unit, spacelike vectors $\bold K, \bold L$, $\bold S$, with components
\begin{equation}
K_\alpha=(0,B,0,0); \quad  L_\alpha=\left(0,0,\frac{\sqrt{A^2B^2r^2+G^2}}{A},0\right),
\label{7}
\end{equation}
\begin{equation}
 L^\alpha=\left(\frac{G}{A\sqrt{A^2B^2r^2+G^2}},0,\frac{A}{\sqrt{A^2B^2r^2+G^2}},0\right),
\label{3n}
\end{equation}
\begin{equation}
 S_\alpha=(0,0,0,C),
\label{3nb}
\end{equation}
satisfying  the following relations:
\begin{equation}
V_{\alpha} V^{\alpha}=-K^{\alpha} K_{\alpha}=-L^{\alpha} L_{\alpha}=-S^{\alpha} S_{\alpha}=-1,
\label{4n}
\end{equation}
\begin{equation}
V_{\alpha} K^{\alpha}=V^{\alpha} L_{\alpha}=V^{\alpha} S_{\alpha}=K^{\alpha} L_{\alpha}=K^{\alpha} S_{\alpha}=S^{\alpha} L_{\alpha}=0.
\label{5n}
\end{equation}

For the energy density and the isotropic pressure, we have
\begin{equation}
\mu=T_{\alpha \beta}V^\alpha V^\beta,\qquad P=\frac{1}{3}h^{\alpha \beta}T_{\alpha \beta},
\label{eisp}
\end{equation}
where
\begin{equation}
h^\alpha_{\beta}=\delta ^\alpha_{\beta}+V^\alpha V_{\beta},
\label{vel5}
\end{equation}
whereas the anisotropic tensor  may be  expressed through three scalar functions defined as (see \cite{1}):

\begin{eqnarray}
 \Pi _{KL}=K^\alpha L^\beta T_{\alpha \beta} 
, \quad , \label{7P}
\end{eqnarray}

\begin{equation}
\Pi_I=(2K^{\alpha} K^{\beta} -L^{\alpha} L^{\beta}-S^{\alpha} S^{\beta}) T_{\alpha \beta},
\label{2n}
\end{equation}
\begin{equation}
\Pi_{II}=(2L^{\alpha} L^{\beta} -S^{\alpha} S^{\beta}-K^{\alpha} K^{\beta}) T_{\alpha \beta}.
\label{2nbis}
\end{equation}

Finally, we may write the heat flux vector in terms of two scalar functions:
\begin{equation}
q_\mu=q_IK_\mu+q_{II} L_\mu,
\label{qn1}
\end{equation}
or, in coordinate components (see \cite{1})

\begin{equation}
q^\mu=\left(\frac{q_{II} G}{A \sqrt{A^2B^2r^2+G^2}}, \frac{q_I}{B}, \frac{Aq_{II}}{\sqrt{A^2B^2r^2+G^2}}, 0\right),\label{q}
\end{equation}
\begin{equation}
 q_\mu=\left(0, B q_I, \frac{\sqrt{A^2B^2r^2+G^2}q_{II}}{A}, 0\right).
\label{qn}
\end{equation}
Of course, all the above quantities depend,  in general, on $t, r, \theta$.

For the  kinematical variables  we obtain   (see  \cite{1}):

For the four acceleration
\begin{equation}
a_\alpha=V^\beta V_{\alpha;\beta}=a_I K_\alpha+a_{II}L_\alpha,
\label{a1n}
\end{equation}
with
\begin{equation}
a_I= \frac {A^\prime }{AB };\quad a_{II}=\frac{A}{\sqrt{A^2B^2r^2+G^2}}\left[\frac {A_{,\theta}}{A}+\frac {G}{A^2}\left(\frac{\dot G}{G}-\frac{\dot A}{A}\right)\right],
\label{acc}
\end{equation}
 where the dot  and the prime denote derivatives with respect to $t$ and $r$ respectively. 

For the expansion scalar
\begin{eqnarray}
\Theta&=&V^\alpha_{;\alpha}=\frac{1}{A}\left(\frac{2 \dot B}{B}+\frac{\dot C}{C}\right) \nonumber\\
&+&\frac{G^2}{A\left(A^2 B^2 r^2 + G^2\right)}\left(-\frac{\dot A}{A}-\frac{\dot B}{B}+\frac{\dot G}{G}\right).
\label{theta}
\end{eqnarray}

Next, the shear tensor
\begin{equation}
\sigma_{\alpha \beta}= V_{(\alpha;\beta)}+a_{(\alpha}
V_{\beta)}-\frac{1}{3}\Theta h_{\alpha \beta}, \label{acc}
\end{equation}

 may be  defined through two scalar functions, as:

\begin{eqnarray}
\sigma _{\alpha \beta}=\frac{1}{3}(2\sigma _I+\sigma_{II}) \left(K_\alpha
K_\beta-\frac{1}{3}h_{\alpha \beta}\right)\nonumber \\+\frac{1}{3}(2\sigma _{II}+\sigma_I) \left(L_\alpha
L_\beta-\frac{1}{3}h_{\alpha \beta}\right).\label{sigmaT}
\end{eqnarray}
The above scalars may be written in terms of the metric functions and their derivatives as (see \cite{1}):
\begin{eqnarray}
\sigma_{I}&=&\frac{1}{A}\left(\frac{\dot B}{B}-\frac{\dot C}{C}\right)\nonumber \\
&+&\frac{G^2}{A\left(A^2 B^2 r^2 + G^2\right)}\left(\frac{\dot A}{A}+\frac{\dot B}{B}-\frac{\dot G}{G}\right),
 \label{sigmasI}
\end{eqnarray}
\begin{eqnarray}
\sigma_{II}&=&\frac{1}{A}\left(\frac{\dot B}{B}-\frac{\dot C}{C}\right)\nonumber \\
&+&\frac{2 G^2}{A\left(A^2 B^2 r^2 + G^2\right)}\left(-\frac{\dot A}{A}-\frac{\dot B}{B}+\frac{\dot G}{G}\right)
\label{sigmas}.
\end{eqnarray}

Finally,  the vorticity  may be described, either by the vorticity vector  $\omega^\alpha$, or the vorticity tensor $\Omega
^{\beta\mu}$, defined as:
\begin{equation}
\omega_\alpha=\frac{1}{2}\,\eta_{\alpha\beta\mu\nu}\,V^{\beta;\mu}\,V^\nu=\frac{1}{2}\,\eta_{\alpha\beta\mu\nu}\,\Omega
^{\beta\mu}\,V^\nu,\label{vomega}
\end{equation}
where $\Omega_{\alpha\beta}=V_{[\alpha;\beta]}+a_{[\alpha}
V_{\beta]}$, and $\eta_{\alpha\beta\mu\nu}$ denote  the Levi-Civita tensor; we find a single component different from zero,  producing:

\begin{equation}
\Omega_{\alpha\beta}=\Omega (L_\alpha K_\beta -L_\beta
K_{\alpha}),\label{omegaT}
\end{equation}
and
\begin{equation}
\omega _\alpha =-\Omega S_\alpha.
\end{equation}
with the scalar function $\Omega$ given by
\begin{equation}
\Omega =\frac{G(\frac{G^\prime}{G}-\frac{2A^\prime}{A})}{2B\sqrt{A^2B^2r^2+G^2}}.
\label{no}
\end{equation}

\subsection{The electric and magnetic part of the Weyl tensor and the super--Poynting vector}
Let us now introduce the electric ($E_{\alpha\beta}$) and magnetic ($H_{\alpha\beta}$) parts of the Weyl tensor ( $C_{\alpha \beta
\gamma\delta}$),  defined as usual by
\begin{eqnarray}
E_{\alpha \beta}&=&C_{\alpha\nu\beta\delta}V^\nu V^\delta,\nonumber\\
H_{\alpha\beta}&=&\frac{1}{2}\eta_{\alpha \nu \epsilon
\rho}C^{\quad \epsilon\rho}_{\beta \delta}V^\nu
V^\delta\,.\label{EH}
\end{eqnarray}

The electric part of the Weyl tensor has only three independent non-vanishing components, whereas only two components define the magnetic part. Thus  we may  write these tensors, in terms of three ($\mathcal{E}_I, \mathcal{E}_{II}, \mathcal{E}_{KL}$) and two ($H_1, H_2$), scalar functions, respectively as.

\begin{widetext}
\begin{equation}
E_{\alpha\beta}=\frac{1}{3}(2\mathcal{E}_I+\mathcal{E}_{II})\left (K_\alpha
K_\beta-\frac{1}{3}h_{\alpha \beta}\right) +\frac{1}{3}(2\mathcal{E}_{II}+\mathcal{E}_{I}) \left(L_\alpha
L_\beta-\frac{1}{3}h_{\alpha \beta}\right)+\mathcal{E}_{KL} (K_\alpha
L_\beta+K_\beta L_\alpha), \label{E'}
\end{equation}
\end{widetext}
\noindent

and
\begin{equation}
H_{\alpha\beta}=H_1(S_\alpha K_\beta+S_\beta
K_\alpha)+H_2(S_\alpha L_\beta+S_\beta L_\alpha)\label{H'}.
\end{equation}

Also, from  the Riemann tensor we may define  three tensors $Y_{\alpha\beta}$, $X_{\alpha\beta}$ and
$Z_{\alpha\beta}$ as

\begin{equation}
Y_{\alpha \beta}=R_{\alpha \nu \beta \delta}V^\nu V^\delta,
\label{Y}
\end{equation}
\begin{equation}
X_{\alpha \beta}=\frac{1}{2}\eta_{\alpha\nu}^{\quad \epsilon
\rho}R^\star_{\epsilon \rho \beta \delta}V^\nu V^\delta,\label{X}
\end{equation}
and
\begin{equation}
Z_{\alpha\beta}=\frac{1}{2}\epsilon_{\alpha \epsilon \rho}R^{\quad
\epsilon\rho}_{ \delta \beta} V^\delta,\label{Z}
\end{equation}
 where $R^\star _{\alpha \beta \nu
\delta}=\frac{1}{2}\eta_{\epsilon\rho\nu\delta}R_{\alpha
\beta}^{\quad \epsilon \rho}$  and $\epsilon _{\alpha \beta \rho}=\eta_{\nu
\alpha \beta \rho}V^\nu$.

The above tensors in turn, may be  decomposed, so that each of them is described through  four scalar functions known as structure scalars \cite{sc}. These are (see \cite{1} for details)
\begin{eqnarray}
Y_T&=&4\pi(\mu+3P), \qquad X_T=8\pi \mu, \label{ortc1}\\
Y_I&=&\mathcal{E}_I-4\pi \Pi_I, \qquad X_I=-\mathcal{E}_I-4\pi \Pi_I
\label{ortc2}\\
Y_{II}&=&\mathcal{E}_{II}-4\pi \Pi_{II},\qquad X_{II}=-\mathcal{E}_{II}-4\pi \Pi_{II}, \label{YY}\\
Y_{KL}&=&\mathcal{E}_{KL}-4\pi \Pi_{KL}, \quad X_{KL}=-\mathcal{E}_{KL}-4\pi \Pi_{KL}\nonumber\\.\label{KL}
\end{eqnarray}

\begin{widetext}
\begin{equation}
Z_I=(H_1-4\pi q_{II});\quad Z_{II}=(H_1+4\pi  q_{II}); \quad Z_{III}=(H_2-4\pi q_I); \quad  Z_{IV}=(H_2+4\pi q_I). \label{Z2}
\end{equation}
\end{widetext}
From the above tensors, we may define  the super--Poynting
vector  by
\begin{equation}
P_\alpha = \epsilon_{\alpha \beta \gamma}\left(Y^\gamma_\delta
Z^{\beta \delta} - X^\gamma_\delta Z^{\delta\beta}\right),
\label{SPdef}
\end{equation}
 in our case,  we may write:
\begin{equation}
 P_\alpha=P_I K_\alpha+P_{II} L_\alpha,\label{SP}
\end{equation}

with
\begin{widetext}
\begin{eqnarray}
P_I =
\frac{2H_2}{3}(2{\cal E}_{II}+{\cal E}_I)+2H_1{\cal E}_{KL}+ \frac{32\pi^2 q_I}{3}\left[3(\mu+P)+\Pi_I\right] 
+32\pi^2 q_{II}\Pi_{KL} ,\nonumber
\\
P_{II}=-\frac{2H_1}{3}(2{\cal E}_{I}+{\cal E}_{II})-2H_2{\cal E}_{KL}
+ \frac{32\pi^2 q_{II}}{3}\left[3(\mu+P)+\Pi_{II}\right]+32\pi^2q_I\Pi_{KL} . \label{SPP}
\end{eqnarray}
\end{widetext}

In the theory of  the super--Poynting vector, a state of gravitational radiation is associated to a  non--vanishing component of the latter (see \cite{11p, 12p, 14p}). This is in agreement with the established link between the super--Poynting vector and the news functions \cite{5p}, in the context of the Bondi--Sachs approach \cite{7, 8}. 

We can identify two different contributions in (\ref{SPP}). On the one hand we have contributions from the  heat transport process. These are in principle independent of the magnetic part of the Weyl tensor, which explains why they  remain in the spherically symmetric limit.  Next we have contributions related to the gravitational radiation. These require, both, the electric and the magnetic part of the Weyl tensor to be different from zero.

\section{The null fluid outside the source}
As it was mentioned in the Introduction, if the source produces gravitational radiation, then an entropy production factor should be present in its hydrodynamic description. This is so, because as it has been discussed before in \cite{7}, \cite{hetaln},  gravitational radiation is an irreversible process (once causality condition is impossed), and therefore this fact should show up in the equation of state of the source. The obvious consequence of the presence of these dissipative processes within the source, is the existence of incoherent radiation, outside the source. 

Thus, we assume that outside the source there is a  null fluid, which due to the symmetry constraints, has  to be described by the energy momentum tensor of the form
\begin{equation}
T_{\alpha \beta}=\lambda l_\alpha l_\beta+\epsilon n_\alpha n_\beta,
\label{1nf}
\end{equation}
where $\lambda$ and $\epsilon$ are two functions of $t,r,\theta$ related with the energy density of the null radiation in either direction $\bold l$ and $\bold n$, and these  two null vectors are given by
\begin{equation}
l^\alpha=\left(\frac{1}{A}, \frac{1}{B}, 0, 0\right)\quad n^\alpha=\left(\frac{1}{A}, 0, -\frac{G+\sqrt{A^2B^2r^2+G^2}}{AB^2r^2}, 0\right),
\label{2nf}
\end{equation}
or
\begin{equation}
l_\alpha=\left(-A, B, \frac{G}{A}, 0\right),
\label{3nf}
\end{equation}

\begin{widetext}
\begin{equation}
n_\alpha=\left[-A-\frac{G}{AB^2r^2}(G+\sqrt{A^2B^2r^2+G^2}), 0, -\frac{\sqrt{A^2B^2r^2+G^2}}{A}, 0\right].
\label{4nf}
\end{equation}
\end{widetext}

We can now expres the vectors $\bold l$ and $\bold n$ in terms of the tetrad  vectors $\bold V, \bold K, \bold L, \bold S$.

Thus we find

\begin{equation}
l_\alpha=V_\alpha+K_\alpha,
\label{5nf}
\end{equation}

and 

\begin{equation}
n_\alpha=\alpha V_\alpha+\gamma L_\alpha,
\label{6nf}
\end{equation}
where

\begin{equation}
\alpha\equiv 1+\frac{G}{A^2B^2r^2}(G+\sqrt{A^2B^2r^2+G^2}),
\label{7nf}
\end{equation}
and 
\begin{equation}
\gamma\equiv -1-\frac{\alpha G}{\sqrt{A^2B^2r^2+G^2}},
\label{8nf}
\end{equation}
implying $\alpha=-\gamma$.

Now comparing (\ref{6bis}) with (\ref{1nf}), we find the following equivalence between different physical variables:
\begin{eqnarray}
q_I&=&\lambda; \quad q_{II}=-\alpha^2  \epsilon; \quad P=\frac{1}{3}(\lambda+\epsilon \alpha^2);\quad \Pi_{KL}=0;\quad \nonumber \\ \Pi_I&=&2\lambda -\epsilon\alpha^2;\quad \Pi_{II}=2\epsilon\alpha^2-\lambda; \quad \mu=\lambda+\epsilon \alpha^2.
\label{9nfp}
\end{eqnarray}

Observe that we have the equation of state $P=\frac{\mu}{3}$ corresponding to a pure radiation gas, as should be for a null fluid distribution.

Now, the interest of (\ref{9nfp}) resides in the fact, that we can apply all the formalism developped in \cite{1}, to the study of our null fluid, just changing the variables according to the relationships  indicated above. However three important differences with the interior (source) case, must be pointed out, namely:
\begin{itemize}
\item Since we are now considering the outside of the source,  the center of the fluid distribution is excluded from the space--time under study, accordingly no regular conditions at the center have to be impossed.
\item Since our source is assumed to be bounded, we have to imposse asymptotic conditions at spatial infinity. In particular we shall assume that our line element approaches asymptotically the Minkowski metric.
\item At the boundary of the source, appropriate junction conditions (Darmois) must be impossed to avoid the presence of shells. Even though, for the study presented here, such conditions will no be used explicitely, they have to be taken into account for any specific global model, describing the space--time ouside and inside the fluid distribution.
\end{itemize}

As  should be obvious, there is not a unique space--time corresponding to our null fluid distribution (as is the case for the spherically symmetric situation), we have instead an infinite number of possible solutions.  Thus our main purpose here is not to provide specific solutions to this case, but rather to bring out some specific aspects of the problem. More precisely, we would like to exhibit the role played by the vorticity in the properties of the null fluid. Thus we shall consider the vorticity--free case.

\subsection{The case without vorticity $G=0$}
Thus, let us assume the vanishing of the vorticity ($G=0$) then $\alpha=-\gamma=1$, which implies  because of (\ref{9nfp})
\begin{eqnarray}
q_I&=&\lambda; \quad q_{II}=- \epsilon; \quad P=\frac{1}{3}(\lambda+\epsilon);\quad \Pi_{KL}=0;\quad \nonumber \\ \Pi_I&=&2\lambda -\epsilon;\quad \Pi_{II}=2\epsilon-\lambda; \quad \mu=\lambda+\epsilon.
\label{9nf}
\end{eqnarray}
Also in this case 
\begin{eqnarray}
\sigma_I=\sigma_{II}=\bar \sigma&=&\frac{1}{A}\left(\frac{\dot B}{B}-\frac{\dot C}{C}\right),\nonumber \\
\Theta&=&\frac{1}{A}\left(\frac{2\dot B}{B}+\frac{\dot C}{C}\right).
\label{1vf}
\end{eqnarray}
First, let us recall that  as $r\rightarrow \infty$ we must recover the Minkowski spacetime, implying that we can write, at least sufficiently far from the source:

\begin{eqnarray}
A(t,r, \theta)&=&\sum_{n \geq0}\frac{A^{(n)}(t, \theta)}{ r^{n}};\quad B(t,r,\theta)=\sum_{n \geq0}\frac{B^{(n)}(t,\theta)}{ r^{n}}\nonumber \\&&C(t,r, \theta)=\sum_{n \geq-1}\frac{C^{(n)}(t, \theta)}{ r^{n}},
\label{es15}
\end{eqnarray}
where $A^{(0)}=B^{(0)}=1$, $C^{(-1)}=\sin \theta$,  $C^{(0)}$=0.

Also we can write

\begin{equation}
\lambda(t,r,\theta)=\sum_{n \geq 1}\frac{\lambda^{(n)}(t,\theta)}{ r^{n}};\quad \epsilon(t,r,\theta)=\sum_{n \geq 1}\frac{\epsilon^{(n)}(t,\theta)}{ r^{n}}.
\label{es17}
\end{equation}
We shall first prove that in the case $G=0$, either $\lambda=\epsilon=0$ or we have the spherically symmetric situation.

Indeed, using (\ref{es15}) and (\ref{es17}) in (\ref{b6}), (\ref{b7}) and (\ref{b76}), we obtain at order $O(r^{-1})$ and $O(r^{-3})$,
\begin{equation}
\epsilon^{(1)}=\epsilon^{(2)}=0.
\label{es18vfg}
\end{equation}

Next, combining (\ref{b7bis}) with (\ref{b76}) we may write
\begin{widetext}
\begin{eqnarray}
-\frac{\dot \epsilon}{A}+\frac{\epsilon_{,\theta}}{Br}+\frac{\epsilon}{B}\left(\frac{A^\prime}{A}-\frac{(Br)^\prime}{Br}\right)+\frac{\epsilon}{Br}\left(\frac{2A_{,\theta}}{A}+\frac{B_{,\theta}}{B}+\frac{C_{\theta}}{C}\right)-\frac{\epsilon}{A}\left(\frac{3\dot B}{B}+\frac{\dot C}{C}\right)=0.
\label{b7bbis}
\end{eqnarray}
\end{widetext}

Feeding back  (\ref{es15}) and (\ref{es17}) into (\ref{b7bbis}) and using (\ref{es18}), we find at order  $O(r^{-3})$ and $O(r^{-4})$, that
\begin{equation}
\dot \epsilon^{(3)}=0,\qquad \dot \epsilon^{(4)}=\epsilon^{(3)}_{,\theta}+\epsilon^{(3)}(\cot \theta-1-3\dot B^{(1)}).
\label{es19vfg}
\end{equation}

Now, for any physically meaningful radiating process, we must demand $\epsilon$  to vanish out of a finite time interval, implying that $\dot \epsilon^{(3)}=0$ $\Rightarrow$ $ \epsilon^{(3)}=0$, which in turn, using (\ref{es19vfg}), implies  $\dot \epsilon^{(4)}=0$ $\Rightarrow$  $\epsilon^{(4)}=0$.

 Following this line of arguments at all possible orders, it is found (MATHEMATICA was used for this purpose)
\begin{equation}
\dot \epsilon^{(n)}=0,  \forall  n \Rightarrow \epsilon^{(n)}=0 \Rightarrow \epsilon=0.
\label{es20vfg}
\end{equation}

Then from (\ref{b76}) we see that either $\lambda=0$, or we have a spherically symmetric system with $A_{,\theta}=B_{,\theta}=0$.

Let us first consider the case $\lambda=\epsilon=0$.  Then we obtain from (\ref{b6}) and (\ref{b7})
\begin{equation}
\bar \sigma^\prime C_{,\theta}-\bar \sigma_{,\theta} C^\prime=0,
\label{es16vf}
\end{equation}
and
\begin{equation}
\Theta^\prime C_{,\theta}-\Theta_{,\theta} C^\prime=0,
\label{es17vf}
\end{equation}
whose combination produces, using (\ref{1vf})
\begin{equation}
\left(\frac{\dot C}{AC}\right)^\prime C_{,\theta}-\left(\frac{\dot C}{AC}\right)_{,\theta} C^\prime=0.
\label{es18vf}
\end{equation}

Feeding back (\ref{es15}) into the above equation, it is easy to prove that for any $n \geq 1$
\begin{equation}
\dot C^{(n)}=\frac{f(t)}{\sin^n\theta},
\label{es19vf}
\end{equation}
where $f$ is an arbitrary function of its argument. 

Then from regularity conditions on the symmetry axis we must put $f=0$, which implies
\begin{equation}
\dot C=0.
\label{es20vf}
\end{equation}

The above result implies, because of (\ref{1vf})
\begin{equation}
\Theta=2\bar \sigma.
\label{es21vf}
\end{equation}

Then the integration of (\ref{b6}), using the above equation and (\ref{es16vf}), produces 
\begin{equation}
\bar \sigma=g(t) C,
\label{es22vf}
\end{equation}
where $g$ is an arbitrary function of its argument.

Finally, feeding back the above expression into (\ref{b8}) and (\ref{b9}) we obtain
\begin{equation}
H_1=-\frac{g(t) C_{,\theta}}{Br};\qquad H_2=\frac{g(t) C^\prime}{B}.
\label{es23vf}
\end{equation}

However, since we must impose the asymptotic condition $H_1,H_2\Rightarrow 0$ as $r\rightarrow \infty$, then we must put $g(t)=0$, implying that the metric is static. We shall further discuss on this issue in the last section.

Next we shall consider the spherically symmetric case.

\subsection{The spherically symmetric limit}

It is instructive to reproduce the spherically symmetric case (Vaidya), in the context of the formalism considered here. 

In the spherically symmetric case the following  conditions  apply:
\begin{equation}
H_1=H_2=\bar \sigma=a_{II}=q_{II}=\epsilon=\Pi_{KL}=A_{,\theta}=B_{,\theta}=0,
\label{es1}
\end{equation}
and 
\begin{equation}
C=Br\sin\theta;\qquad q_I=\lambda;\qquad  a_I=\frac{A^\prime}{AB},
\label{es2}
\end{equation}
implying
\begin{eqnarray}
&&\Theta=\frac{3\dot B}{AB};\qquad \Pi_I=2\lambda;\qquad \Pi_{II}=-\lambda;\qquad \mu=\lambda, \nonumber\\ &&P=\frac{\lambda}{3};\quad 2\Pi_{II}+\Pi_I= {\cal E}_{KL}=Y_{KL}=X_{KL}=0.
\label{es3}
\end{eqnarray}

Using the conditions above in (\ref{b6}) we obtain
\begin{equation}
\left(\frac{1}{A}\frac{\dot B}{B}\right)^\prime=4\pi B\lambda,
\label{es4}
\end{equation}

whereas (\ref{b2}) and (\ref{b4}) produce  respectively

\begin{eqnarray}
2a_I^\prime+a_I\left(\frac{2A^\prime}{A}-\frac{2(Br)^\prime}{Br}\right)-B{\cal E}_I+8\pi B\lambda=0,
\label{es5}
\end{eqnarray}

\begin{eqnarray}
a_I^\prime+a_I\left(\frac{A^\prime}{A}-\frac{(Br)^\prime}{Br}\right)+B{\cal E}_{II}+4\pi B\lambda=0.
\label{es6}
\end{eqnarray}
From the two equations above, it follows at once
\begin{equation}
2{\cal E}_{II}+{\cal E}_I=0\Rightarrow 2X_{II}+X_I=2Y_{II}+Y_I=0.
\label{es7}
\end{equation}
Next, (\ref{b1}) reads
\begin{eqnarray}
\frac{1}{A}\dot \Theta+\frac{1}{3}\Theta^2-\frac{1}{B}\left[a_I^\prime+a_I\left(\frac{A^\prime}{A}+\frac{2(Br)^\prime}{Br}\right)\right]+8\pi\lambda=0,\nonumber \\
\label{es8}
\end{eqnarray}
whereas (\ref{b3}, \ref{b5}, \ref{b7}, \ref{b8}, \ref{b9}) become identities.

Let us now turn to the conservation laws. We obtain from (\ref{b6c})
\begin{eqnarray}
\frac{\dot \lambda}{A}+\frac{4\lambda \dot B}{AB}+\frac{1}{B}\left[\lambda^\prime+2\lambda\left(\frac{A^\prime}{A}+\frac{(Br)^\prime}{Br}\right)\right]=0,
\label{es9}
\end{eqnarray}
whereas (\ref{b7bbb}) produces the same result as (\ref{es9}), and (\ref{b7bis}) (\ref{b76}) become identities.

Next, we obtain from (\ref{b10}), (\ref{b12}), (\ref{b13}) and (\ref{b14}), respectively
\begin{eqnarray}
\frac{\dot {\cal E}_I}{3A}+\frac{4\pi \dot \lambda}{A}+\frac{{\cal E}_I \Theta}{3}=-\frac{8\pi}{3} \lambda\Theta- \frac{4\pi}{B}\left(\lambda^\prime+\frac{2A^\prime \lambda}{A}\right),\nonumber \\
\label{es11}
\end{eqnarray}

\begin{eqnarray}
\frac{\dot {\cal E}_{II}}{3A}+\frac{{\cal E}_{I I}\Theta}{3}=-\frac{4\pi}{3} \lambda \Theta-\frac{4\pi \lambda (Br)^\prime}{B^2r},
\label{es12}
\end{eqnarray}

\begin{eqnarray}
-\frac{\dot {\cal E}_I+\dot{\cal E}_{II}}{3A}-\frac{{(\cal E}_I +{\cal E}_{II})\Theta}{3}=-\frac{4\pi}{3} \lambda\Theta-\frac{4\pi \lambda}{B}\frac{(Br)^\prime}{Br},\nonumber \\
\label{es13}
\end{eqnarray}

\begin{eqnarray}
\frac{1}{3B}\left[{\cal E}_I^\prime+3{\cal E}_I\frac{(Br)^\prime}{Br}\right]+8\pi \lambda\frac{(Br)^\prime}{B^2r}=-\frac{8\pi \lambda}{3} \Theta,
\label{es14}
\end{eqnarray}

whereas (\ref{b11}), (\ref{b15}), (\ref{b16}), and (\ref{b17}) become identities.

It is a simple matter to check that (\ref{b18})  becomes identical to (\ref{es14}), and that (\ref{es12}) and (\ref{es13}) are equivalent.

We can now  determine, asymptotically,  the spacetime (Vaidya) by the  iterative procedure scketched below.

First, let us notice that  equations  (\ref{es4}, \ref{es8}, \ref{es9}),  may be written respectively,  as

\begin{equation}
\frac{\dot B^{\prime}}{AB}-\frac{A^\prime \dot B}{A^2B}-\frac{\dot BB^\prime}{AB^2}=4\pi B \lambda,
\label{es18}
\end{equation}

\begin{equation}
\frac{3 \ddot B}{A^2B}-\frac{3\dot A \dot B}{A^3B}-\frac{A^{\prime \prime}}{AB^2}-\frac{A^{\prime}B^\prime}{AB^3}-\frac{2A^{\prime}}{AB^2r}=-8\pi  \lambda,
\label{es19}
\end{equation}

\begin{equation}
\frac{ \dot \lambda}{A}+\frac{4\lambda \dot B}{AB}+\frac{\lambda^{\prime}}{B}+\frac{2A^{\prime} \lambda}{AB}+\frac{2\lambda B^{\prime}}{B^2}+\frac{2\lambda}{Br}=0.
\label{es20}
\end{equation}

Using (\ref{es15}), and  evaluating (\ref{es18}) at the order $O(r^{-1})$, $O(r^{-2})$ and $O(r^{-3})$, and (\ref{es20}) at order $O(r^{-3})$, we find that  (obviously in the spherically symmetric case the coefficcients $A^{(n)}$,  $B^{(n)}$, and  $\lambda^{(n)}$, do not depend on $\theta$)
\begin{eqnarray}
\lambda^{(1)}=0;\qquad \lambda^{(2)}=-\frac{\dot B^{(1)}}{4\pi};\quad \dot \lambda^{(3)}=-4\lambda^{(2)}\dot B^{(1)}.
\label{es21}
\end{eqnarray}
 
Next, evaluating (\ref{es20}) at order $O(r^{-2})$, it follows that
\begin{equation}
\dot  \lambda^{(2)}=0 \Rightarrow \lambda^{(2)}=constant.
\label{es23}
\end{equation}

This last result together with (\ref{es21}) produces 
\begin{equation}
B^{(1)}=-4\pi\lambda^{(2)}t+ constant,
\label{es24}
\end{equation}

and

\begin{equation}
\lambda^{(3)}=16\pi(\lambda^{(2)})^{2}t+constant.
\label{es25}
\end{equation}

Next, the order $O(r^{-3})$  in (\ref{es18}) produces 

\begin{eqnarray}
 -2\dot B^{(2)}=12\pi B^{(1)}\lambda^{(2)}+8\pi A^{(1)}\lambda^{(2)}+ 4\pi \lambda^{(3)}.
\label{es22}
\end{eqnarray}

From the above equation we cannot obtain the time dependence of $B^{(2)}$ since we do not know $ A^{(1)}$.

So, let us turn to (\ref{es19}), at the highest order( $O(r^{-1}$)) we find

\begin{equation}
\ddot B^{(1)} =0,
\label{es26}
\end{equation}
a known result, whereas from the order $O(r^{-2})$ it follows
\begin{equation}
12\pi \lambda^{(2)}\dot A^{(1)}+3\ddot B^{(2)}=-8\pi \lambda^{(2)}.
\label{es27}
\end{equation} 
Now, taking $t$-derivative of (\ref{es22}),
solving for $\ddot B^{(2)}$ and feeding back into (\ref{es27}) we obtain 
\begin{equation}
 \lambda^{(2)}=\frac{1}{3\pi}.
\label{es28}
\end{equation} 

In other words we still need a function of time in order to determine 
the time dependence of  $A^{(1)}$ and  $B^{(2)}$.

So, let us look for the next orders in (\ref{es18}, \ref{es19}, \ref{es20}).
Thus, from the order $O(r^{-4})$ in (\ref{es18}), $O(r^{-3})$ in (\ref{es19}) and $O(r^{-4})$ in (\ref{es20}) we obtain respectively:
\begin{eqnarray}
4\pi\lambda^{(4)}&=&-\dot B^{(1)}\left[3(A^{(1)})^2+6(B^{(1)})^2+5A^{(1)}B^{(1)}-3A^{(2)}-4B^{(2)}\right]\nonumber \\&+&\dot B^{(2)}\left(5B^{(1)}+3A^{(1)}\right)-3\dot B^{(3)},
\label{es29}
\end{eqnarray} 
\begin{widetext}
\begin{eqnarray}
-8\pi\lambda^{(3)}=-3\dot B^{(2)}\dot A^{(1)}+\dot B^{(1)}\left[-3\dot A^{(2)}+3\dot  A^{(1)}B^{(1)}+9\dot A^{(1)}A^{(1)}\right]+3\ddot B^{(3)}-3\ddot B^{(2)}\left(2A^{(1)}+B^{(1)}\right),
\label{es30}
\end{eqnarray} 
\end{widetext}
 \begin{widetext}
\begin{eqnarray}
\dot \lambda^{(4)}-\dot \lambda^{(3)}A^{(1)}=\lambda^{(2)}\left(4\dot B^{(1)}A^{(1)}+4\dot B^{(1)}B^{(1)}-4\dot B^{(2)}+2A^{(1)}+2B^{(1)}\right)+\lambda^{(3)}(1-4\dot B^{(1)}).
\label{es31}
\end{eqnarray} 
\end{widetext}

Once $B^{(2)}$ or $A^{(1)}$ have been determined, we can find $\lambda^{(4)}$ from (\ref{es31}). 

Next, taking the time derivative of (\ref{es29}) and combining with (\ref{es30}), we obtain
\begin{widetext}
\begin{eqnarray}
4\pi\dot \lambda^{(4)}-8\pi\lambda^{(3)}=-\dot B^{(1)}\left(-3A^{(1)}\dot A^{(1)}+12B^{(1)}\dot B^{(1)}+2B^{(1)}\dot A^{(1)}+5A^{(1)}\dot B^{(1)}-9\dot B^{(2)}\right)+\ddot B^{(2)}(2B^{(1)}-3A^{(1)}).
\label{es32}
\end{eqnarray} 
\end{widetext}
Thus we have no further information about  the time dependence of $A^{(2)}$ or $B^{(3)}$, which implies that we have to provide the time dependence of either one of them.
Following this procedure {\it ad nauseam} we obtain, as expected,  that the metric is obtained up to an  arbitrary function of $t$ and $r$. Thus,  whereas the Vaidya metric has an extremely simple form in null coordinates, in the present approach it is only possible  to construct it asymptotically,  as a series expansion.

\section{Conclusions}
 We have tackled the problem of describing the outer space--time of axially symmetric sources of gravitational radiation, based in  the  Bondi conjecture \cite{7} (confirmed in \cite{hetaln}), according to which the process of gravitational radiation is an irreversible one, and therefore must entail dissipative proceses within the source. 

The ensuing consequence of this, is that there should be an  incoherent radiation (null fluid)  at the outside of the source, produced by those dissipative processes. Keeping this fact in mind, we should remark that the Bondi--Sachs metric \cite{7}, \cite{8}, should be regarded as an approximation to the space--time outside the source, when the null fluid produced by the dissipative processes is neglected.

 Starting with the description of this null fluid, we apply the formalism developped in \cite{1}, to study some of the properties of such a null fluid. 

As the main result of our study we find that the absence of vorticity implies that the exterior spacetime is either static or spherically symmetric (Vaidya). Reinforcing thereby the fundamental role of vorticity in any process involving production of gravitational radiation, already stressed in \cite{vorh}, \cite{mr}. 

There exists still the possibility of  the non--radiative, non--static solutions considered by Bondi in \cite{7} (see also \cite{4nu}), which correspond to the case $G\neq 0$, $H_1=H_2=0$.  Indeed, in this case the Bondi's news function vanishes  \cite{4nu}, and therefore the system does not radiate gravitational waves (this is also evident from (\ref{SPP})), even though it may be  time dependent. In particular, it  can be shown  that  the mass, the ``dipole'' and the ``quadrupole''
moments (as defined in \cite{7}) correspond to a static situation. However, the time dependence might enter through coefficients of higher order in the metric, giving rise to what Bondi calls ``non--natural, non--radiative moving systems''. In this latter case the  three first moments are time  independent,but the systema allows for time dependence of higher moments (see also \cite{B1}, \cite{B2}). As unlikely as this situation may be from the physical point of view, it cannot be ruled out.

It must be kept in mind, that all along our discussion we have restricted ourselves, to physically meaningful situations, where the source is bounded and the radiation process takes place during a  finite time interval. Obviously if we relax either of these conditions, another cases might appear, even though they would be deprived of physical relevance.

Finally, we have indicated how to recover (at least asymptotically) the Vaidya metric. Unlike the null coordinates, our coordinates do not allow for a simple expression for the corresponding line element.

\begin{acknowledgments}
 L.H and J.O. acknowledge financial support from the Spanish Ministry of Science and Innovation (grant FIS2009-07238) and  Fondo Europeo de Desarrollo Regional (FEDER) (grant FIS2015-65140-P) (MINECO/FEDER).
\end{acknowledgments}

\appendix 
\section{Some basic equations}
In what follows, we shall present the main equations of the formalism, specialized for the case with $G=0$ (which of course includes as a particular case the spherically symmetric situation).

Thus, from B1, B2, B3 and B4 in \cite{1} we get respectively:
\begin{widetext}
\begin{eqnarray}
\frac{1}{A}\dot \Theta+\frac{1}{3}\Theta^2+\frac{2}{3}\bar \sigma^2-\frac{1}{B}\left[a_I^\prime+a_I\left(\frac{A^\prime}{A}+\frac{(Br)^\prime}{Br}+\frac{C^\prime}{C}\right)\right]-\frac{1}{Br}\left[a_{II,\theta}+a_{II}\left(\frac{A_{,\theta}}{A}+\frac{B_{,\theta}}{B}+\frac{C_{,\theta}}{C}\right)\right]+8\pi(\lambda+\epsilon)=0,\nonumber\\
\label{b1}
\end{eqnarray}
\end{widetext}
\begin{widetext}
\begin{eqnarray}
&&\frac{1}{A}\dot {\bar \sigma}+\frac{1}{A^2}\left[\left(\frac{\dot B}{B}\right)^2-\left(\frac{\dot C}{C}\right)^2\right]-\frac{1}{B}\left[2a_I^\prime+a_I\left(\frac{2A^\prime}{A}-\frac{(Br)^\prime}{Br}-\frac{C^\prime}{C}\right)\right]+\frac{1}{Br}\left[a_{II,\theta}+a_{II}\left(\frac{A_{,\theta}}{A}-\frac{2B_{,\theta}}{B}+\frac{C_{,\theta}}{C}\right)\right]\nonumber \\&+&{\cal E}_I-4\pi(2\lambda-\epsilon)=0,
\label{b2}
\end{eqnarray}
\end{widetext}
\begin{widetext}
\begin{eqnarray}
\frac{1}{B}\left[a_{II}^\prime-a_{II}\frac{(Br)^\prime}{Br}\right]+\frac{1}{Br}\left[a_{I,\theta}+a_{I}\left(\frac{2A_{,\theta}}{A}-\frac{B_{,\theta}}{B}\right)\right]-2{\cal E}_{KL}=0,
\label{b3}
\end{eqnarray}
\end{widetext}
\begin{widetext}
\begin{eqnarray}
&&\frac{1}{A}\dot {\bar \sigma}+\frac{1}{A^2}\left[\left(\frac{\dot B}{B}\right)^2-\left(\frac{\dot C}{C}\right)^2\right]+\frac{1}{B}\left[a_I^\prime+a_I\left(\frac{A^\prime}{A}-\frac{2(Br)^\prime}{Br}+\frac{C^\prime}{C}\right)\right]-\frac{1}{Br}\left[2a_{II,\theta}+a_{II}\left(\frac{2A_{,\theta}}{A}-\frac{B_{,\theta}}{B}-\frac{C_{,\theta}}{C}\right)\right]\nonumber \\&+&{\cal E}_{II}-4\pi(2\epsilon-\lambda)=0.
\label{b4}
\end{eqnarray}
\end{widetext}
From (\ref{b2}) and (\ref{b4}) we obtain
\begin{widetext}
\begin{eqnarray}
\frac{1}{B}\left[-a_I^\prime+a_I\left(-\frac{A^\prime}{A}+\frac{(Br)^\prime}{Br}\right)\right]+\frac{1}{Br}\left[a_{II,\theta}+a_{II}\left(\frac{A_{,\theta}}{A}-\frac{B_{,\theta}}{B}\right)\right]=\frac{{\cal E}_{II}-{\cal E}_{I}}{3}-4\pi(\epsilon-\lambda).
\label{b34}
\end{eqnarray}
\end{widetext}

Next, from B5 in \cite{1}, we obtain

\begin{widetext}
\begin{eqnarray}
-\frac{1}{B}\left[a_{II}^\prime+a_{II}\frac{(Br)^\prime}{Br}\right]+\frac{1}{Br}\left[a_{I,\theta}+a_{I}\frac{B_{,\theta}}{B}\right]=0,
\label{b5}
\end{eqnarray}
\end{widetext}
which combined with (\ref{b3}) produces
\begin{equation}
{\cal E}_{KL}=\frac{1}{B^2r}\left[\frac{A^\prime_{,\theta}}{A}-\frac{A^\prime B_{,\theta}}{AB}-\frac{A_{,\theta}(Br)^\prime}{ABr}\right].
\label{b35}
\end{equation}

Next, from B6, B7, B8 and B9 in \cite{1} we get respectively
\begin{equation}
\frac{1}{3}\left(2\Theta-\bar  \sigma\right)^\prime-\bar \sigma\frac{C^\prime}{C}=8\pi B\lambda,
\label{b6}
\end{equation}
\begin{equation}
\frac{1}{3}\left(2\Theta-\bar  \sigma\right)_{,\theta}-\bar \sigma\frac{C_{,\theta}}{C}=-8\pi Br\epsilon,
\label{b7}
\end{equation}

\begin{equation}
H_1=-\frac{1}{2Br}\left(\bar  \sigma_{,\theta}+\bar \sigma\frac{C_{,\theta}}{C}\right),
\label{b8}
\end{equation}

\begin{equation}
H_2=\frac{1}{2B}\left(\bar  \sigma^\prime+\bar \sigma\frac{C^\prime}{C}\right).
\label{b9}
\end{equation}

We have next the conservation laws (eqs.A6, A7 in \cite{1}), which read
\begin{widetext}
\begin{eqnarray}
\frac{1}{A}{(\dot \lambda+\dot \epsilon)}+\frac{(\lambda+\epsilon)}{A}\left(\frac{3\dot B}{B}+\frac{\dot C}{C}\right)+\frac{1}{B}\left[\lambda^\prime+\lambda\left(\frac{2A^\prime}{A}+\frac{(Br)^\prime}{Br}+\frac{C^\prime}{C}\right)\right]-\frac{1}{Br}\left[\epsilon_{,\theta}+\epsilon\left(\frac{2A_{,\theta}}{A}+\frac{B_{,\theta}}{B}+\frac{C_{,\theta}}{C}\right)\right]=0,\nonumber\\
\label{b6c}
\end{eqnarray}
\end{widetext}
\begin{widetext}
\begin{eqnarray}
\frac{\dot \lambda}{A}+\frac{\lambda^\prime}{B}+\frac{\lambda}{B}\left[\frac{2A^\prime}{A}+\frac{(Br)^\prime}{Br}+\frac{C^\prime}{C}\right]+\frac{\epsilon}{B}\left[\frac{A^\prime}{A}-\frac{(Br)^\prime}{Br}\right]+\frac{\lambda}{A}\left(\frac{3\dot B}{B}+\frac{\dot C}{C}\right)=0,
\label{b7bbb}
\end{eqnarray}
\end{widetext}
and
\begin{widetext}
\begin{eqnarray}
-\frac{\dot \epsilon}{A}+\frac{\epsilon_{,\theta}}{Br}+\frac{\lambda}{Br}\left(\frac{A_{,\theta}}{A}-\frac{B_{,\theta}}{B}\right)+\frac{\epsilon}{Br}\left(\frac{2A_{,\theta}}{A}+\frac{B_{,\theta}}{B}+\frac{C_{\theta}}{C}\right)-\frac{\epsilon}{A}\left(\frac{3\dot B}{B}+\frac{\dot C}{C}\right)=0,
\label{b7bis}
\end{eqnarray}
\end{widetext}
the combination of the last three equations produces,

\begin{eqnarray}
\frac{\lambda}{Br}\left(\frac{A_{,\theta}}{A}-\frac{B_{,\theta}}{B}\right)=\frac{\epsilon}{B}\left[\frac{A^\prime}{A}-\frac{(Br)^\prime}{Br}\right].
\label{b76}
\end{eqnarray}

Finally, from B10, B11, B12, B13, B14, B15, B16, B17 and B18 in \cite{1}, we obtain respectively:
\begin{widetext}
\begin{eqnarray}
\frac{\dot {\cal E}_I}{3A}&+&\frac{4\pi \dot \lambda}{A}+\frac{{\cal E}_I \Theta}{3}+\frac{{\cal E}_{II} \bar \sigma}{3}-\frac{1}{Br}\left[H_{1,\theta}+H_1\left(\frac{2A_{,\theta}}{A}+\frac{C_{,\theta}}{C}\right)\right]+\frac{H_2}{B}\left[\frac{(Br)^\prime}{Br}-\frac{C^\prime}{C}\right]=-\frac{4\pi}{3} (2\lambda+\epsilon)(\bar \sigma+\theta)\nonumber \\&-&\frac{4\pi}{B}\left(\lambda^\prime+\frac{2A^\prime \lambda}{A}\right)+\frac{4\pi \epsilon B_{,\theta}}{B^2r},
\label{b10}
\end{eqnarray}

\begin{eqnarray}
\frac{\dot {\cal E}_{KL}}{A}&+&{\cal E}_{KL} (\Theta-\bar \sigma)+\frac{1}{2B}\left[H_1^\prime+H_1\left(\frac{2A^\prime}{A}-\frac{(Br)^\prime}{Br}+\frac{2C^\prime}{C}\right)\right]-\frac{1}{2Br}\left[H_{2,\theta}+H_2\left(\frac{2A_{,\theta}}{A}-\frac{B_{,\theta}}{B}+\frac{2C_{,\theta}}{C}\right)\right]\nonumber \\&=&\frac{2\pi}{B} \left[\epsilon^\prime+\epsilon\left(\frac{2A^\prime}{A}-\frac{(Br)^\prime}{Br}\right)\right]-\frac{2\pi}{Br}\left[\lambda_{,\theta}+\lambda\left(\frac{2A_{,\theta}}{A} -\frac{B_{,\theta}}{B}\right)\right],
\label{b11}
\end{eqnarray}

\begin{eqnarray}
\frac{\dot {\cal E}_{II}}{3A}&+&\frac{4\pi \dot \epsilon}{A}+\frac{{\cal E}_{I I}\Theta}{3}+\frac{{\cal E}_{I} \bar \sigma}{3}+\frac{1}{B}\left[H^\prime_2+H_2\left(\frac{2A^\prime}{A}+\frac{C^\prime}{C}\right)\right]-\frac{H_1}{Br}\left(\frac{B_{,\theta}}{B}-\frac{C_{,\theta}}{C}\right)=-\frac{4\pi}{3} (\lambda+2\epsilon)(\bar \sigma+\theta)\nonumber \\&+&\frac{4\pi}{Br}\left(\epsilon_{,\theta}+\frac{2A_{,\theta} \epsilon}{A}\right)-\frac{4\pi \lambda (Br)^\prime}{B^2r},
\label{b12}
\end{eqnarray}
\begin{eqnarray}
-\frac{\dot {\cal E}_I+\dot{\cal E}_{II}}{3A}&-&\frac{{(\cal E}_I +{\cal E}_{II})(\Theta+\bar \sigma)}{3}+\frac{1}{Br}\left[H_{1,\theta}+H_1\left(\frac{2A_{,\theta}}{A}+\frac{B_{,\theta}}{B}\right)\right]-\frac{1}{B}\left[H_2^\prime+H_2\left(\frac{2A^\prime}{A}+\frac{(Br)^\prime}{Br}\right)\right]\nonumber \\&=&\frac{4\pi}{3} (\lambda+\epsilon)(2\bar \sigma-\Theta)-\frac{4\pi \lambda}{B}\frac{C^\prime}{C}+\frac{4\pi \epsilon C_{,\theta}}{BCr},
\label{b13}
\end{eqnarray}
\end{widetext}

\begin{widetext}
\begin{eqnarray}
&&\frac{1}{3B}\left[{\cal E}_I^\prime+{\cal E}_I\left(\frac{(Br)^\prime}{Br}+\frac{2C^\prime}{C}\right)\right]-\frac{{\cal E}_{II}}{3B}\left[\frac{(Br)^\prime}{Br}-\frac{C^\prime}{C}\right]+\frac{1}{Br}\left[{\cal E}_{KL,\theta}+{\cal E}_{KL}\left(\frac{2B_{,\theta}}{B}+\frac{C_{,\theta}}{C}\right)\right]-\frac{4\pi}{B}\left[\epsilon^\prime+\frac{\epsilon (Br)^\prime}{Br}\right]\nonumber\\&+&\frac{4\pi \lambda}{B}\left[\frac{(Br)^\prime}{Br}+\frac{C^\prime}{C}\right]-H_2 \bar \sigma=-\frac{4\pi \lambda}{3}(2\Theta-\bar \sigma),
\label{b14}
\end{eqnarray}

\begin{eqnarray}
&&\frac{1}{3Br}\left[{\cal E}_{II,\theta}+{\cal E}_{II}\left(\frac{B_{,\theta}}{B}+\frac{2C_{,\theta}}{C}\right)\right]-\frac{{\cal E}_{I}}{3Br}\left(\frac{B_{,\theta}}{B}-\frac{C_{,\theta}}{C}\right)+\frac{1}{B}\left[{\cal E}^\prime_{KL}+{\cal E}_{KL}\left(\frac{2(Br)^\prime}{Br}+\frac{C^\prime}{C}\right)\right]-\frac{4\pi}{Br}\left(\lambda_{,\theta}+\frac{\lambda B_{,\theta}}{B}\right)\nonumber\\&+&\frac{4\pi \epsilon}{Br}\left(\frac{B_{,\theta}}{B}+\frac{C_{,\theta}}{C}\right)+H_1 \bar \sigma=\frac{4\pi \epsilon}{3}(2\Theta-\bar \sigma),
\label{b15}
\end{eqnarray}

\begin{eqnarray}
\frac{1}{B}\left[H_1^\prime+H_1\left(\frac{(Br)^\prime}{Br}+\frac{2C^\prime}{C}\right)\right]+\frac{1}{Br}\left[H_{2,\theta}+H_2\left(\frac{B_{,\theta}}{B}+\frac{2C_{,\theta}}{C}\right)\right]=\frac{4\pi}{B}\left[\epsilon^\prime+\frac{\epsilon (Br)^\prime}{Br}\right]+\frac{4\pi}{Br}\left(\lambda_{,\theta}+\frac{\lambda B_{,\theta}}{B}\right),\nonumber\\
\label{b16}
\end{eqnarray}
\begin{eqnarray}
&&-\frac{1}{3Br}\left[{\cal E}_{I,\theta}+{\cal E}_{I}\left(\frac{2A_{,\theta}}{A}+\frac{C_{,\theta}}{C}\right)\right]-\frac{1}{3Br}\left[{\cal E}_{II,\theta}+{\cal E}_{II}\left(\frac{A_{,\theta}}{A}+\frac{2C_{,\theta}}{C}\right)\right]+\frac{{\cal E}_{KL}}{B}\left(\frac{A^\prime}{A}-\frac{C^\prime}{C}\right)+\frac{4\pi}{Br}\left(\lambda_{,\theta}+\frac{\lambda B_{,\theta}}{B}\right)\nonumber\\&-&\frac{4\pi \epsilon}{Br}\frac{B_{,\theta}}{B}+\frac{\dot H_1}{A} +H_1\Theta +\frac{4\pi \epsilon}{3}(\Theta+\bar \sigma)=0,
\label{b17}
\end{eqnarray}
\begin{eqnarray}
&&\frac{1}{3B}\left[{\cal E}^\prime_{I}+{\cal E}_{I}\left(\frac{A^\prime}{A}+\frac{2C^\prime}{C}\right)\right]+\frac{1}{3B}\left[{\cal E}^\prime_{II}+{\cal E}_{II}\left(\frac{2A^\prime}{A}+\frac{C^\prime}{C}\right)\right]-\frac{{\cal E}_{KL}}{Br}\left(\frac{A_{,\theta}}{A}-\frac{C_{,\theta}}{C}\right)-\frac{4\pi}{B}\left[\epsilon^\prime+\frac{\epsilon (Br)^\prime}{Br}\right]\nonumber\\&+&\frac{4\pi \lambda}{B}\frac{(Br)^\prime}{Br}+\frac{\dot H_2}{A} +H_2\Theta +\frac{4\pi \lambda}{3}(\Theta+\bar \sigma)=0.
\label{b18}
\end{eqnarray}
\end{widetext}

\end{document}